\let\originalLaTeXLabel\label
\documentclass[10pt,aps,prl,twocolumn,showpacs,preprintnumbers,amsmath,amssymb,
floatfix,superscriptaddress]{revtex4-1}

\usepackage{ulem}
\usepackage{graphics}
\usepackage{graphicx}
\usepackage{amsmath}
\usepackage{setspace}
\usepackage{braket}
\usepackage{epstopdf}
\usepackage{comment}
\let\label\originalLaTeXLabel
\makeatletter
\let\ltx@label\label
\makeatother
\usepackage{hyperref}
\usepackage{bm}
\usepackage{xfrac}
\usepackage{blindtext}
\usepackage{mathrsfs}
\usepackage{xcolor}
\usepackage[version=4]{mhchem}

\newcommand{\affA}{Department of Physics, Stony Brook University, Stony Brook, New York 11794, USA}

\begin{document}

\title{Sympathetic Cooling of Ionic Fullerenes with Ultracold Atoms to Millikelvin Temperatures}
\author{Saajid Chowdhury}\affiliation{\affA}
\author{Zhongqi Liang}\affiliation{\affA}
\author{Yu Wang}\affiliation{\affA}
\author{Jes\'us P\'erez-R\'{i}os}\affiliation{\affA}
\date{\today}

\begin{abstract}
Cooling complex molecular ions to millikelvin temperatures remains a major challenge because their intricate internal structures prevent laser cooling. Here, we propose to use an ultracold gas of Cd atoms to efficiently cool C$_{60}^+$ ions to average kinetic energies of approximately 20~mK, following an initial precooling stage with He buffer gas. Our approach is based on full molecular dynamics simulations using accurate ion-atom interaction potentials derived from high-level electronic structure methods. Furthermore, using a Monte Carlo approach, we predict the formation of C$_{60}^+$ Coulomb crystals supported solely by collisions with the buffer gas. The probability of reaching these cold states shows a pronounced dependence on the number of trapped ions, revealing the competition between neutral-atom cooling and many-ion dynamics in radio-frequency traps.  
\end{abstract}

\maketitle

Fullerene ions are fascinating molecular systems characterized by highly symmetric structures and rich spectroscopic properties. Among them, C$_{60}^+$ is of particular interest because it has been identified as a carrier of diffuse interstellar bands (DIBs)~\cite{C60plus_1,C60plus_2,C60plus_review,C60plus_review2}. This identification was made possible only by gas-phase spectroscopy in a cryogenic environment~\cite{DIBs_C60plus}. Further cooling these complex ions to the mK regime could open new avenues to study their interactions with neutral systems and possibly help elucidate their formation mechanisms under conditions similar to those of the interstellar medium~\cite{Gerlich,Cold_ions_ISM,Lewandowski}. 

Cooling and controlling complex molecular ions remains challenging because their intricate internal structures generally preclude conventional laser cooling~\cite{Nguyen_2011,Vogelius_2002}. An alternative is sympathetic cooling, i.e., cooling by contact with a colder system, in which complex ions thermalize to a lower temperature through elastic collisions. Sympathetic cooling of complex ions comes in two flavors: (1) sympathetic cooling with cold ions and (2) with neutrals. In the first scenario, the idea is to use cold atomic ions that are continuously laser-cooled to dissipate the complex ions' kinetic energy via Coulomb interaction~\cite{Bowe_1999,Stefan_2014,Blythe_2205,C60plus_sympathetic_cooling_with_ions,Michael,Larson_sympathetic_cooling}. However, as more ions are added, the Coulomb interaction pushes ions far from the center of the trap and RF heating becomes significant. In the second scenario, the idea is to bring the ions into contact with a cold gas of neutrals so atom-ion collisions control thermalization \cite{Hudson_sympahetic_cooling_neutrals,10.1039/b820977d,Hudson_blue_sky,Cetina_2012,Holtkemeir_2016}. These collisions can be highly reactive, so the neutral gas must be selected with caution. Furthermore, trapped ions interacting with cold neutrals can form atom-ion complexes that induce heating, depending on the ion-to-atom mass ratio, putting further constraints on the neutral system under consideration.

In this Letter, we propose sympathetic cooling of C$_{60}^+$ ions with cold Cd atoms as a route to obtain millikelvin complex molecular ions. Our full molecular dynamics simulations, using potential energy curves from high-level electronic structure methods, indicate that an ionic fullerene will thermalize to about 20~mK after a pre-cooling stage with He atoms, without Cd-tagging. Furthermore, using a Monte Carlo approach, we predict that these ions form linear Coulomb crystals~\cite{Morigi2026IonCoulombCrystals} just through collisions with the neutral gas. As the number of ions increases, more trajectories form hot clouds which do not cool, while others become cold and highly localized. Indeed, the competition between sympathetic cooling and chaotic ion RF heating (which hinders crystallization beyond a certain initial rms cloud radius~\cite{Blumel1989Chaos}) has been shown to produce a blue-sky bifurcation in the energy of trapped ions~\cite{Hudson_blue_sky}, highlighting the challenges of reaching cold states through neutral-atom collisions alone.

%, reflecting a competition between a slowly evolving hot cloud and an already cold localized state

%%% This is great %%%
%There is a competition between the Brownian kicks that collide head on with the ion to genuinely slow it down, competing with lateral Brownian kicks which deflect the ions away from their equilibrium positions where they want to go, giving the other ions an opportunity to further deflect them and induce micromotion heating. This is not a problem with Doppler cooling, which slows down the ions without deflecting them away from equilibrium positions between other ions. 

\begin{figure*}
    \centering
    \includegraphics[width=\linewidth]{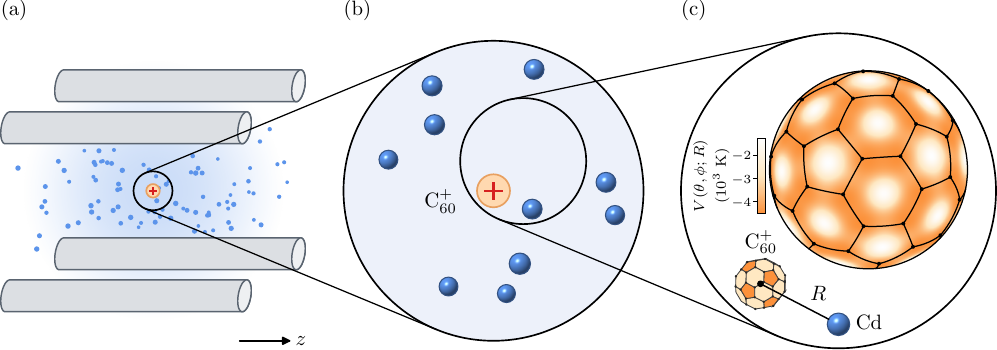}
    \caption{Schematic of the cooling of C$_{60}^+$ by cadmium atoms. (a) Ion in a quadrupole trap surrounded by a cloud of cold cadmium atoms. (b) Zoom-in on the full MD atom-launching sphere of radius $r=(10/(4\pi n/3))^{1/3}\approx 2.1$~$\mu$m surrounding the ion and the 10 simulated Cd atoms. (c) Fit of our calculation of the C$_{60}^+$-Cd interaction as a function of direction for a fixed C$_{60}^+$-Cd distance $R=6.0$~{\AA}. The interaction potential shown ranges from -4500~K to -1250~K.}
    \label{fig:fig1}
\end{figure*}

First, we consider a single C$_{60}^+$ ion in a time-dependent rf field of a quadrupole trap, where $\Omega_\textrm{rf}=2\pi\times 1.433$ MHz, $a_i=(+1.8,-2.9,+1.1)\times 10^{-4}$, and $q_i=(-0.24,+0.24,0)$. Here, the DC stability parameter $a_i\propto 4eU_\textrm{DC}/m\Omega_\textrm{rf}^2$ characterizing the static field strength and the RF stability parameter $q_i\propto 2eV_\textrm{RF}/m\Omega_\textrm{rf}^2$ characterizing the alternating field strength describe the motion of an isolated trapped ion via the Mathieu equation $(4/\Omega_\textrm{rf}^2)d^2x_i/dt^2+(a_i-2q_i\cos\Omega_\textrm{rf}t)x_i=0$. The ion interacts with a buffer gas of atoms, He or Cd, whose interaction is dominated by the isotropic potentials $V_{00}(r)$ which in the long range are $-\alpha/2r^4$ where $\alpha$ is the polarizability of the atom in atomic units. A schematic representation of the system is shown in Fig.~\ref{fig:fig1}(a). We include full time-dependent atom-ion dynamics classically with the nearest 10 atoms in our full MD approach, shown schematically in Fig.~\ref{fig:fig1}(b), and use Newtonian kinematics to compute instantaneous momentum transfers from atoms in our MC approach. 

%In our model, C$_{60}^+$ is treated as a point particle which interacts with surrounding He or Cd atoms via our calculated isotropic potentials $V_{00}(r)$ which in the long range are $-\alpha/2r^4$ where $\alpha$ is the polarizability of the atom in atomic units. The ion dynamics is treated classically with molecular dynamics in the time-dependent rf field of a quadrupole trap, where $\Omega_\textrm{rf}=2\pi\times 1.433$ MHz, $a_i=(+1.8,-2.9,+1.1)\times 10^{-4}$, and $q_i=(-0.24,+0.24,0)$. The system is shown in Fig.~\ref{fig:fig1}(a). We include full time-dependent atom-ion dynamics classically with the nearest 10 atoms in our full MD approach (shown schematically in Fig.~\ref{fig:fig1}(b)), and use Newtonian kinematics to compute instantaneous momentum transfers from atoms in our MC approach. 

The interaction energy of C$_{60}^+$ with He and Cd are computed using the density functional theory (DFT) framework in Gaussian 16~\cite{Gaussian}. Following the approaches of Ref.~\cite{C60_elastic}, we employ the $\omega$B97XD functional~\cite{WB97XD} and the 6-31G(d,p) basis set for C$_{60}^+$--He. For C$_{60}^+$--Cd, we instead use the def2-SV(P) basis set for all atoms. The interaction energies are also corrected for the basis-set superposition error (BSSE) by the counterpoise method. The geometry of C$_{60}^+$ is optimized separately for each system and held fixed for single-point calculations with He or Cd at different positions. Following Ref.~\cite{C60_elastic}, we expand the PESs in Racah-normalized spherical harmonics $C_{lm}(\theta,\phi)$ with radial coefficients $V_{lm}(r)$ determined by least-squares fitting $V(\boldsymbol{r})=\sum_{l,m} V_{lm}(r) (C_{l,m}(\theta,\phi)+C_{l,-m}(\theta,\phi))/2$. We retain the terms of $l=0,6,10,12,16,18,20$ and $m=5n$ ($0\leq m\leq l$) used for icosahedral C$_{60}$ there and also add to the list $l=2,4$ to allow for the Jahn--Teller distortion of C$_{60}^+$~\footnote{Including more terms, such as $l=8$, produces a negligible improvement in the fit's accuracy but significantly increases its instability. }. An interpolative fit to the calculated potential energy values is shown in Fig.~\ref{fig:fig1}(c). 

\begin{figure*}
    \centering
    \includegraphics[width=\linewidth]{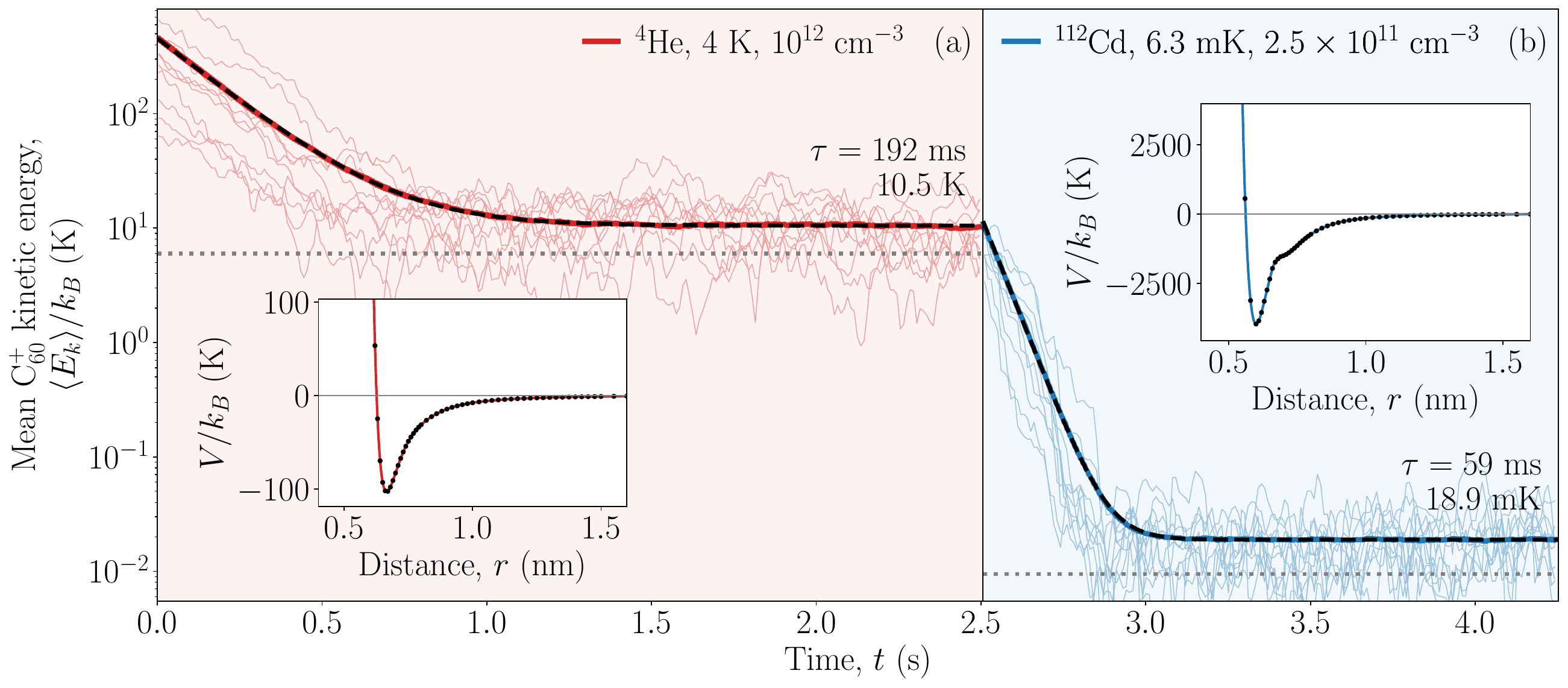}
    \caption{Average kinetic energy of C$_{60}^+$ as a function of simulated time. (a) The first phase cools C$_{60}^+$ from an initial average kinetic energy 450~K to 10.5~K via buffer gas cooling of $n=10^{12}$ cm$^{-3}$ helium at 4~K. (b) The second phase cools C$_{60}^+$ starting from its final conditions from the first phase, to a final average kinetic energy of 18.9~mK, via a buffer gas of $n=2.5\times 10^{11}$ cm$^{-3}$ cadmium at 6.3~mK. We run and average over hundreds of independent trajectories with random initial conditions: 576 trajectories for the first phase and 2439 trajectories for the second. The main curve is the kinetic energy of the C$_{60}^+$ as a function of time, averaged over all independent trajectories and over 1~ms windows. The faint curves are 10 individual trajectories, averaged over 10~ms windows. The insets show the calculated PES isotropic component, and a cubic spline fit connected via a third-degree Hermite polynomial to a $-C_4/r^4$ tail which was used in the simulations. We fit each main thermalization curve to $E(t)\sim A+Be^{-t/\tau}$, shown by the black dotted curves, giving characteristic cooling times of 192~ms in helium and 59~ms in cadmium. The horizontal dotted lines show the atom bath average kinetic energy, $\frac{3}{2}k_BT_\textrm{bath}$.}
    \label{fig:fig2}
\end{figure*}

The isotropic components $V_{00}(r)$ are extracted for subsequent dynamical simulations. For C$_{60}^+$--He, due to the high ionization energy of He, the positive charge remains localized on the fullerene cage even at short separations, producing an isotropic potential that varies smoothly with $r$, as shown in the inset of Fig.~\ref{fig:fig2}(a). In contrast, the isotropic component for Cd, as depicted in the inset of Fig.~\ref{fig:fig2}(b), displays much deeper interaction potential and a shoulder in the $0.6$ -- $0.7$~nm region. The shoulder's appearance is due to the onset of charge transfer at intermediate distances, discussed in the end matter.

To simulate the ion interacting with the buffer gas, we calibrate the initial conditions and then run the simulation in two phases: cooling by 4K helium gas, and cooling by 6.3mK cadmium gas, following the results from Ref.\cite{PadillaCastillo2025}. Throughout, we run full MD simulations of an ensemble of thousands of trajectories of a trapped C$_{60}^+$ polyatomic ion interacting with the closest 10 atoms in the gas, following the methodology outlined in Ref.~\cite{chowdhury2026moleculardynamicssimulationssingle}. The simulated region is the sphere of radius $r=(10/(4\pi n/3))^{1/3}$ centered at the instantaneous position of the ion. Whenever an atom leaves the simulated region, a new one with a Maxwell-Boltzmann velocity distribution is spawned at a random position on the surface of the sphere via rejection sampling with acceptance probability proportional to $\max(0,-v_{\textrm{rel},r})$, where $v_{\textrm{rel},r}$ is the radial component of the relative velocity of the atom with respect to the ion. To accurately handle the collision dynamics while simulating efficiently during collisionless periods, we use a fourth/fifth order adaptive timestep Runge Kutta integration method. 

For the calibration stage, we start the C$_{60}^+$ at the origin of the trap with zero initial velocity and run full MD simulations with an artificially hotter bath, such that the average kinetic energy of the ion converges to and hovers around 450~K, adjusting the bath temperature via a negative feedback loop to maintain this average. We ran 2690 independent trajectories for this simulation. Next, for the first cooling phase due to collisions with helium, each trajectory starts with the C$_{60}^+$ at the final position and velocity of a trajectory from the calibration (hence, its initial position and velocity follow the distribution for trapped ions at temperature 300~K), and the bath is composed of helium atoms at temperature 4~K and density $n=10^{12}$ cm$^{-3}$. The simulation is run for a fixed number of timesteps for 576 trajectories, such that the trajectories typically simulate beyond 2.5~s of physical time. We find that C$_{60}^+$ cools to 10.5 K within about 1.5~s, as shown in Fig.~\ref{fig:fig2}(a), while forming almost no complexes. 

In these simulations, we neglect rotational inelastic collisions and describe the ion–atom interaction using its isotropic component, which accounts for approximately 85\% of the full interaction potential at the equilibrium separation. Although this indicates that the isotropic contribution dominates the interaction, anisotropic components may still induce rotational transitions. Indeed, recent calculations predict significant rotational inelastic scattering in C$_{60}$-Ar collisions, although at higher collision energies than those relevant to the present study~\cite{petrov2026quantumscatteringfullerene12c60}. The effects of the anisotropic part of the interaction potential are discussed in the end matter.

For the second cooling phase, when we bring the ion in contact with an ultracold gas of Cd atoms, each trajectory starts with the ion at the final position and velocity of a trajectory from the previous cooling phase, so that the ion's average kinetic energy is initially 10.5~K for this phase. We now simulate a bath of cadmium atoms at temperature 6.3~mK and density $2.5\times 10^{11}$ cm$^{-3}$, which can be loaded into a MOT from a cryogenic buffer gas beam \cite{PadillaCastillo2025}. For 2439 trajectories, the cooling by cadmium is simulated for a fixed number of timesteps which typically lead to at least 1.75~s of simulated time. In the end, the average kinetic energy of the C$_{60}^+$ ion is 18.9~mK, shown in Fig.~\ref{fig:fig2}(b). Cooling is substantially more efficient than during the He stage, consistent with the larger mass of Cd and the stronger C$_{60}^+$-Cd interaction, which can enhance collisional momentum transfer. However, the isotropic component accounts for only approximately 70\% of the total interaction potential at the equilibrium separation, compared to 85\% for C$_{60}^+$-He. This suggests that anisotropic interactions may play a more important role in Cd collisions, potentially enhancing rotational inelastic processes not included in the present simulations.

During the second cooling stage, the Cd atoms are confined in a magneto-optical trap (MOT). Therefore, the ion will be continuously exposed to the laser radiation that keeps the atoms trapped and cold. This laser, in principle, could induce unwanted electronic excitation of C$_{60}^+$. To assess this possibility, we calculate the electronic absorption spectrum of C$_{60}^+$ using time-dependent density functional theory (TDDFT), including 300 excited states and employing the same exchange-correlation functional and basis set used to obtain the ion-atom interaction potentials. Our results indicate non-negligible absorption near the conventional Cd MOT wavelength of 229 nm, suggesting that the cooling light may interfere with the trapped molecular ions. However, Cd atoms can also be trapped using the narrow $^1S_0\rightarrow{}^3P_1$ transition at 326 nm~\cite{Gibble_2024}, where our calculations predict negligible C$_{60}^+$ absorption. This alternative offers a promising route to implementing sympathetic cooling while minimizing unwanted optical excitation of the molecular ions. Moreover, the Doppler temperature associated with this transition is only $1.6~\mu$K, substantially below the 6.3 mK Cd temperature considered in our simulations. Thus, a narrow-line Cd MOT could simultaneously suppress unwanted optical excitation of C$_{60}^+$ and provide a colder atomic bath, potentially enabling sympathetic cooling of fullerene ions to temperatures far below our prediction.

\begin{figure}
    \centering
    \includegraphics[width=\linewidth]{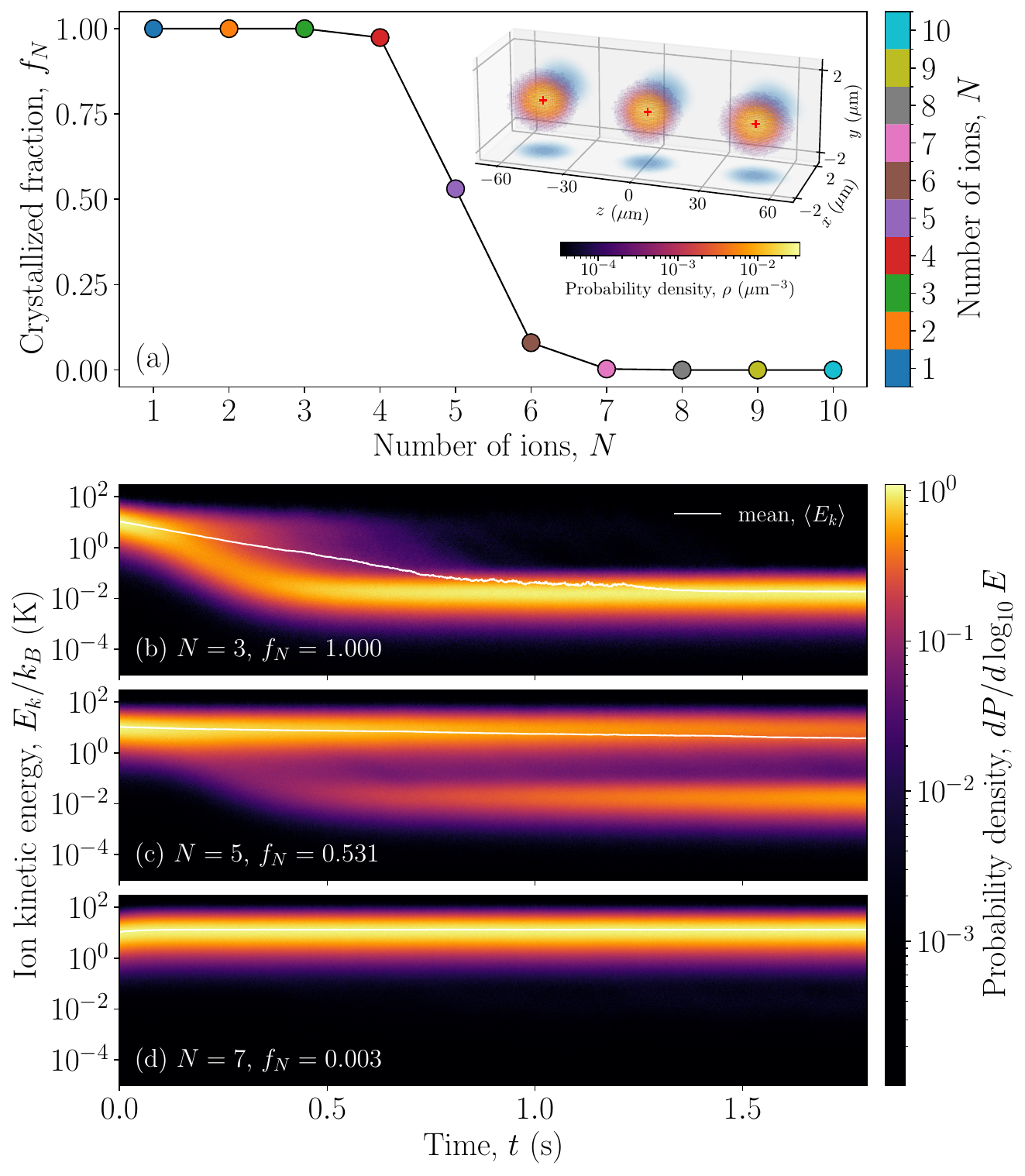}
    \caption{Coulomb crystals of trapped C$_{60}^+$ ions in a buffer gas of $n=2.5\times 10^{11}$ cm$^{-3}$ cadmium at 6.3~mK. (a) Crystallization fraction $f_N$, the fraction of 1000 trajectories simulated up to 1.8~s where ions stop swapping $z$-coordinate-ordering after 1.5~s. The inset shows the probability density of ion positions in a Coulomb crystal with three ions. (b),(c),(d) C$_{60}^+$ kinetic energy distribution over all trajectories over 1~ms averaging windows as a function of simulated time, for $N=3,5,7$.}
    \label{fig:fig3}
\end{figure}

Next, we want to explore the possibility of cooling more than one C$_{60}^+$ ion. To this end, we implement a Monte Carlo (MC) approach where collisions are treated as instantaneous elastic events, occurring at rate $n \sigma_d v_\textrm{rel}$ and kicking the ion isotropically. In particular, we use a fixed timestep symplectic leapfrog integrator to simulate the ion in the trap, and at every timestep we sample an atom's velocity from a Maxwell-Boltzmann distribution and accept a collision with this atom with probability $dP=n \sigma_d(E)v_\textrm{rel}\,dt$ where $E=\frac{1}{2}\mu v_\textrm{rel}^2$ is the collision energy, letting the final relative velocity have magnitude $v_\textrm{rel}$ and uniformly random direction in the lab frame and calculating the final ion's velocity using kinematic momentum and energy conservation if the collision is accepted. The diffusion cross section $\sigma_d(E)$ is obtained using a single channel quantum scattering approach described in the end matter. Furthermore, and more importantly, we find that for a single ion, the average kinetic energy as a function of time for full MD fully agrees with that of MC. This is a huge advantage, because this simulation runs significantly faster than full MD. Thus, we use MC to study Coulomb crystals of C$_{60}^+$ ions in an rf trap. 

In Fig.~\ref{fig:fig3}, we show the results of simulating 1000 trajectories of different numbers of trapped ions, $N=1,\ldots,10$. For each trajectory, every ion feels forces from the trap and from other ions, and each ion independently experiences occasional kicks driven by the MC sampling. In the simulation, the initial positions and velocities of the ions are sampled from the final conditions of the ion from $N$ randomly-selected trajectories from the helium cooling phase, and then we run a fixed number of fixed-size timesteps yielding 1.8 seconds of simulated time. 

What we find is that for $N\leq 4$, the ions typically cool down to about 20~mK of average kinetic energy as shown in Fig.~\ref{fig:fig3}(b), just as for a single ion, and they form a cold line of ions along the $z$-axis. This is because $\omega_z^2/\omega_x^2\approx 0.0038$, meaning that the axial confinement is significantly weaker than the radial confinement~\cite{Schiffer1993PhaseTransitions}. We define the trajectory to be  \textit{crystallized} if from 1.5 seconds to 1.8 seconds, the relative ordering of the $z$-coordinates of the ions stops changing. The rms displacement of the ions about their equilibrium positions, normalized by the nearest-neighbor equilibrium distance, during the time interval 1.5~s to 1.8~s, is on average an order of magnitude smaller for crystallized trajectories than for non-crystallized trajectories, with minimal overlap between the two distributions. For $N\leq 4$, almost 100\% of these trajectories crystallize, indicated by $f_N\approx 1$ in Fig.~\ref{fig:fig3}(a).  

However, for $N=5$, about half of the trajectories' ions cool down to 20~mK and crystallize, while for the other half, the ions often dance around the center of the trap for the full duration of the simulation without cooling down. This non-cooldown is more prevalent for greater numbers of ions $N\geq 6$. The ions tend to spread out to hundreds of microns, forming a hot cloud of ions up to 20~K. In principle, if given enough time, these hot clouds of ions would eventually cool down and crystallize (given the gradual cooling from the high to low energy band in Fig.~\ref{fig:fig3}(c)). However, for larger numbers of ions, it would require combinatorially greater luck for the ions to experience just the right sequence of kicks from atoms to simultaneously cool them down and push them into equilibrium with the surrounding ions. 

Coulomb crystals with much greater numbers of ions can be achieved when the main energy-loss mechanism is laser cooling, because Doppler cooling is like a smooth, velocity-dependent friction which slows the ions down, allowing them to get closer to their equilibrium positions without being deflected or jumping over the barriers between local minima. In our case, where the mechanism is sympathetic cooling, the atoms in the bath cause the ions to experience Brownian motion, which would also slow down an isolated ion over time as shown in Fig.~\ref{fig:fig2}(b). However, for $N\geq 2$, atoms can kick each ion away from its desired equilibrium position between surrounding ions, allowing the other ions to push the kicked ion further away from the trap center so that it experiences more micromotion heating, resetting the cooling process. The more ions there are, the more synchronization is required to cool all ions down simultaneously without any one ion being deflected and heated back up.

%%%% This is a possible route for detection of ionic fullerenes

The experimental detection of cold C$_{60}^+$ ions could be achieved using Doppler recooling thermometry. In this approach, C$_{60}^+$ and Ba$^+$ ions are co-trapped and sympathetically cooled through collisions with ultracold Cd atoms. Once the system reaches a steady state, the Ba$^+$ cooling and repumping lasers are switched on, and the time-dependent fluorescence signal is recorded. Since the Ba$^+$ cooling transitions at 493 and 650 nm are expected to produce negligible electronic excitation of C$_{60}^+$, the fluorescence recovery dynamics can be used to infer the initial motional energy of the Ba$^+$ ions and, through their Coulomb coupling, estimate the translational temperature of the fullerene ions. Furthermore, spatially resolved fluorescence imaging could reveal the presence and positions of the dark C$_{60}^+$ ions through gaps in the fluorescent Ba$^+$ Coulomb crystal.

In this Letter, we have demonstrated a route to sympathetic cooling of complex molecular ions with neutral atoms: our full MD simulations show that trapped fullerene ions, C$_{60}^+$, can be cooled to average kinetic energies of around 20~mK by first pre-cooling with helium at 4~K and then cooling with cadmium at 6.3~mK, with negligible cadmium tagging. Furthermore, our MC calculations show that cadmium cooling enables the formation of linear Coulomb crystals of up to 5-6 fullerene ions. As the number of ions increases, the crystallization probability decreases sharply, with some trajectories reaching cold, spatially localized configurations while others remain in slowly evolving, high-energy clouds. This behavior suggests the emergence of long-lived metastable states arising from the competition between sympathetic cooling and many-ion dynamics in radio-frequency traps. Our results establish a promising route toward preparing ultracold complex molecular ions, opening new opportunities to study low energy ion-neutral collisions and molecular processes relevant for interestellar chemistry. 

\medskip
{\it Acknowledgements}---This work was supported by the U.S. National Science Foundation through CAREER Award No. PHY-2440808. We thank Michael Drewsen and David Leibrandt for insightful discussions on the experimental feasibility of the proposed cooling scheme and the underlying physics.

\bibliographystyle{apsrev4-1}

\newpage
\onecolumngrid
\begin{center}
    \textbf{End Matter}
\end{center}
\twocolumngrid

{\it Potential energy surface}---The grid used to sample the atom's radial and angular positions is constructed based on Ref.~\cite{C60_elastic} and adapted to the newly optimized geometries.

Near the potential minimum, the angularly resolved interaction energies deviate from the isotropic estimate by a root-mean-square error (RMSE) of only 9.02 K. In contrast, the much lower ionization energy of Cd allows significant charge transfer between Cd and the fullerene when it is approximately $0.6$ -- $0.7$ nm from the cage center. The onset of charge transfer is orientation-dependent, as the distance between Cd and the nearest carbon depends on the direction of approach, giving rise to the irregularity in the isotropic potential observed in the inset of Fig.~\ref{fig:fig2}(b). To estimate the possibility of nonadiabatic transitions, we also calculate the first excited state energies for selected angular configurations and, using the Landau--Zener formula, obtaining values that are consistently below 10\% in all cases examined.

{\it MC method}---Our MC method uses the diffusion cross section $\sigma_d(E)$ for the C$_{60}^+$-Cd interaction. To obtain the diffusion cross section, for several different collision energies we use the Numerov method to solve the radial Schrodinger equation for each $l=0,1,\ldots,l_\textrm{max}$ to get the phase shifts $\delta_l$ and then calculate $\sigma_d(E)=\frac{4\pi}{k^2}\sum_{l=0}^{l_\textrm{max}}(l+1)\sin^2(\delta_l-\delta_{l+1})$ where $k=\sqrt{2\mu E}/\hbar$ and $l_\textrm{max}$ is chosen sufficiently large for convergence.

Our MC method has two properties which help demonstrate why it predicts the same average kinetic energy as a function of time as our full MD method for a single ion. The properties are that if hypothetically all the atoms at some instant have the same relative velocity to the ion ${\bf v_\textrm{rel}}={\bf v}_\textrm{ion}-{\bf v}_\textrm{atom}$, then $\frac{d}{dt}\langle\sum_i\Delta p_{i,\parallel}\rangle=-n\mu v_\textrm{rel}^2\sigma_d(E)$ and $\frac{d}{dt}\langle\sum_i|\Delta {\bf p}_i|^2\rangle=2 n\mu^2v_\textrm{rel}^3\sigma_d(E)$, where $i$ indicates the number of collisions so far, and $\Delta p_{i,\parallel}$ is the component of the momentum transfer to the ion during the $i^\textrm{th}$ collision parallel to the relative velocity before the collision ${\bf v_\textrm{rel,i}}$. These are the same expressions as would be obtained for scattering at the correct physical rate $n\sigma_\textrm{el}(E)v_\textrm{rel}$ with the correct distribution of scattering angles. This justifies the MC method, which instead samples effective collisions at a different effective rate $n\sigma_\textrm{d}(E)v_\textrm{rel}$ with isotropic scattering angles. 

To derive these properties, we use the definition of the diffusion cross section in terms of the elastic cross section, $\sigma_d(E)=\int\frac{d\sigma_\textrm{el}}{d\Omega}(1-\cos\theta)\,d\Omega$. The momentum transfer for the $i^\textrm{th}$ elastic collision can be expressed in terms of reduced and relative quantities as $\Delta {\bf p}_i=\mu({\bf v_\textrm{rel,f}}-{\bf v_\textrm{rel,i}})$. One can show via trigonometry that $\Delta p_{i,\parallel}=-\mu v_\textrm{rel}(1-\cos\theta)$ and $|\Delta {\bf p_i}|^2=2\mu^2v_\textrm{rel}^2(1-\cos\theta)$. Supposing hypothetically that all gas is moving with relative velocity ${\bf v_\textrm{rel,i}}$, the momentum transfer per collision, times the rate at which collisions occur into each solid angle element $nv_\textrm{rel}\frac{d\sigma_\textrm{el}}{d\Omega}\,d\Omega$, summed over all possible final angles, gives (taking advantage of the factor $(1-\cos\theta)$ to use the definition of $\sigma_d$), $\frac{d}{dt}\langle\sum_i\Delta p_{i,\parallel}\rangle=-n\mu v_\textrm{rel}^2\sigma_d(E)$. The same procedure applied to $|\Delta {\bf p_i}|^2$ gives $\frac{d}{dt}\langle\sum_i|\Delta {\bf p}_i|^2\rangle=2 n\mu^2v_\textrm{rel}^3\sigma_d(E)$. The angle brackets indicate ensemble averages over many trajectories. To account for the actual gas velocity distribution, these formulas simply get integrated against the Maxwell-Boltzmann velocity distribution. Now, in the case of our MC method, the effective collisions occur into each solid angle element at rate $nv_\textrm{rel}\frac{d\sigma_d}{d\Omega}\,d\Omega$, which is $nv_\textrm{rel}\frac{\sigma_d}{4\pi}\,d\Omega$ due to our assumption of effectively-isotropic scattering. This assumption also gives $\langle1-\cos\theta\rangle=1$, so the average momentum transfer quantities per collision are $\langle\Delta p_{i,\parallel}\rangle=-\mu v_\textrm{rel}$ and $\langle|\Delta {\bf p_i}|^2\rangle=2\mu^2v_\textrm{rel}^2$. Summing over all final angles, which amounts to multiplying these by $nv_\textrm{rel}\sigma_d(E)$, gives exactly the same properties. 

It remains to show that satisfying these two properties implies a particular rate of kinetic energy transfer to the ion, $\frac{d\langle K_\textrm{ion}\rangle}{dt}=-n\mu v_\textrm{rel}\sigma_d(E){\bf v}_\textrm{CM}\cdot{\bf v}_\textrm{rel}$, which helps explain why both our MD and MC methods yield the same cooling time for a single ion. Consider the ion's kinetic energy change during one collision: $\Delta K_\textrm{ion}=\frac{1}{2}m_\textrm{ion}|{\bf v}_\textrm{ion}+\Delta{\bf v}_\textrm{ion}|^2-\frac{1}{2}m_\textrm{ion}v_\textrm{ion}^2={\bf v}_\textrm{ion}\cdot\Delta{\bf p}_\textrm{ion}+\frac{|\Delta p_\textrm{ion}|^2}{2m_\textrm{ion}}$. The ensemble average sum of these over multiple collisions is $\sum_i\langle\Delta K_{\textrm{ion},i}\rangle$, the collisional contribution to the total change in average kinetic energy. We will express the rate of change of this ensemble average sum in terms of the cross section using the properties. Given that the component of the momentum transfer perpendicular to the initial relative velocity has an azimuthally symmetric distribution by symmetry, the first property implies that the instantaneous rate of change of the average momentum vector itself is $\frac{d}{dt}\langle\sum_i\Delta{\bf p}_i\rangle=-n\mu v_\textrm{rel}\sigma_d(E){\bf v_\textrm{rel}}$. The second property once again gives $\frac{d}{dt}\langle\sum_i|\Delta {\bf p}_i|^2\rangle=2 n\mu^2v_\textrm{rel}^3\sigma_d(E)$. Dotting the first property by ${\bf v}_\textrm{ion}$ and dividing the second property by $2m_\textrm{ion}$ gives $\frac{d\langle K_\textrm{ion}\rangle}{dt}=-n\mu v_\textrm{rel}\sigma_d(E)\left({\bf v}_\textrm{ion}\cdot {\bf v}_\textrm{rel}-\frac{\mu}{m_\textrm{ion}}v_\textrm{rel}^2\right)$ which is equivalent to $\frac{d\langle K_\textrm{ion}\rangle}{dt}=-n\mu v_\textrm{rel}\sigma_d(E){\bf v}_\textrm{CM}\cdot{\bf v}_\textrm{rel}$ since ${\bf v}_\textrm{CM}={\bf v}_\textrm{ion}-\frac{\mu}{m_\textrm{ion}}{\bf v}_\textrm{rel}$. This formula, for the special case of a stationary ion and averaged over the actual distribution of atom velocities, is equivalent to Equation A7 in \cite{Jensen2004}. 

\begin{figure}[!t]
    \centering
    \includegraphics[width=\linewidth]{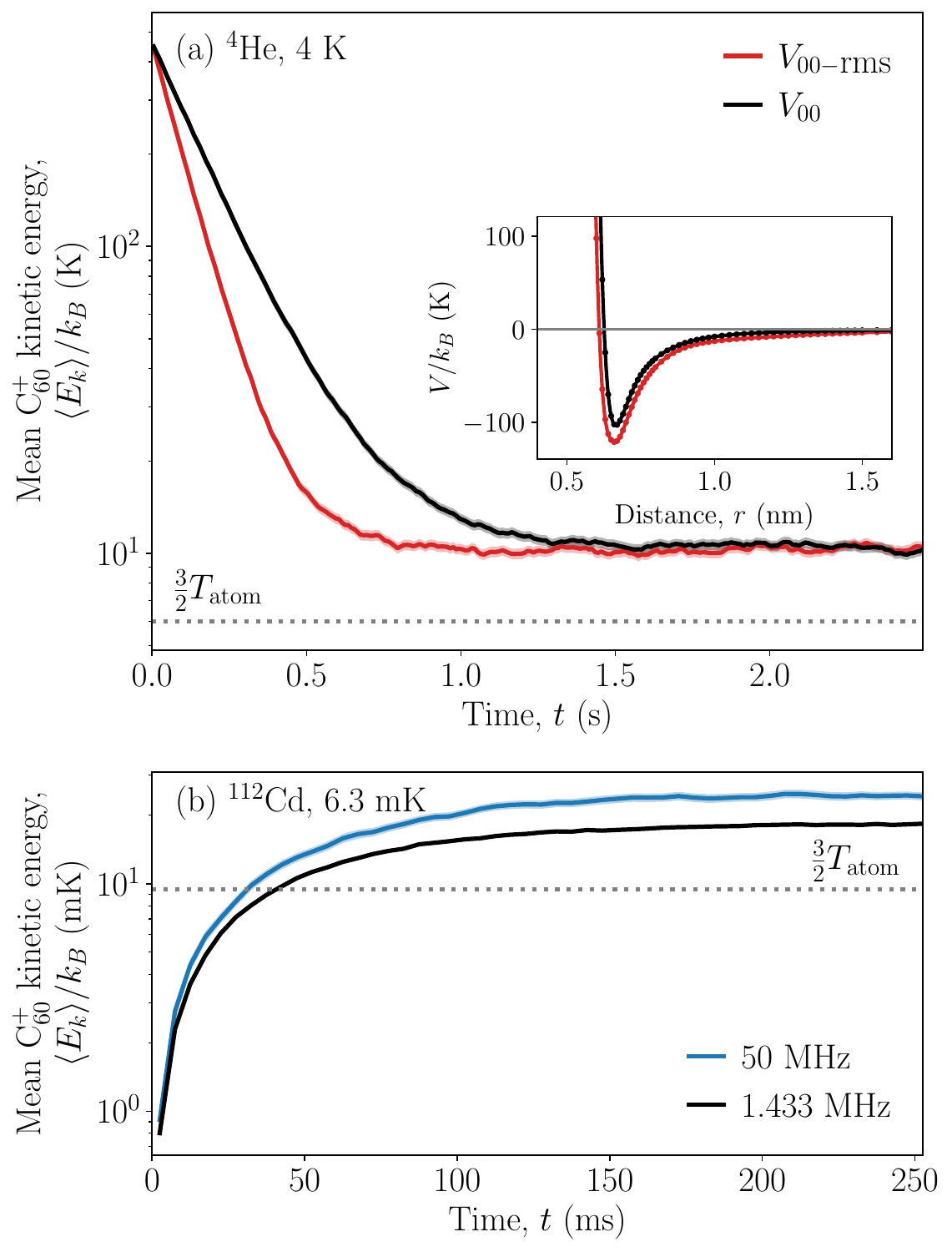}
    \caption{(a) Average C$_{60}^+$ kinetic energy (averaged over all 576 independent trajectories with different random initial conditions, over 1~ms windows) as a function of simulated time for the full MD helium cooling simulation, for two different potentials: the isotropic component $V_{00}(r)$, and an artificial isotropic potential which incorporates the r.m.s. anisotropic effects $V_{00-\textrm{rms}}(r)=V_{00}(r)-\sqrt{\frac{1}{4\pi}\int (V(\theta,\phi;r)^2-V_{00}(r)^2)\,d\Omega}$. (b) Average C$_{60}^+$ kinetic energy as a function of time for cadmium bath with the ion starting at the trap center with zero velocity, under two different rf trap frequencies 1.433~MHz (4583 trajectories) and 50~MHz (576 trajectories).}
    \label{fig:fig4}
\end{figure}

{\it Sensitivity to short range and rf frequency}---The PES for C$_{60}^+$-He is anisotropic, especially in the short range. To check whether this could affect our results for full MD simulations, we run the same full MD simulation with $V_{00-\textrm{rms}}(r)=V_{00}(r)-\sqrt{\frac{1}{4\pi}\int (V(\theta,\phi;r)^2-V_{00}(r)^2)\,d\Omega}$, and compare this to using just $V_{00}(r)$. The two potentials have the same long range (beyond 2.1~nm), but $V_{00-\textrm{rms}}(r)$ is slightly deeper. Fig.~\ref{fig:fig4}(a) shows that both potentials yield the same final ion average kinetic energy, but $V_{00-\textrm{rms}}(r)$ thermalizes about 0.5~s sooner. Hence, although the thermalization rate depends on the details of the potential's short range, the final ion mean kinetic energy does not. 

We also check the dependence of the final average C$_{60}^+$ kinetic energy on the rf frequency of the trap. We compare $\Omega_\textrm{rf}=2\pi\times 1.433$~MHz, as used in Fig.~\ref{fig:fig2}, to a trap with the same $a_i$ and $q_i$ parameters but $\Omega_\textrm{rf}=2\pi\times 50$~MHz. The results in Fig.~\ref{fig:fig4}(b) shows a small increase in the temperature due to the greater micromotion heating from the greater rf frequency; nonetheless, the final ion average kinetic energy remains on the order of tens of mK. 

\end{document}